\documentstyle[11pt,aaspp4,epsf]{article}

\lefthead{Surpi and Harari}
\righthead{Weak lensing and polarization}

\begin{document}

\title{Weak-Lensing by Large-Scale Structure and the \\
Polarization Properties of Distant Radio-Sources}

\author{Gabriela C. Surpi and Diego D. Harari}

\authoremail{surpi@df.uba.ar -- harari@df.uba.ar}

\affil{Departamento de F{\'\i}sica,
Facultad de Ciencias Exactas y Naturales\\
Universidad de Buenos Aires \\
Ciudad Universitaria - Pab. 1,
1428 Buenos Aires, Argentina} 

\begin{abstract}

We estimate the effects of weak lensing by large-scale 
density inhomogeneities and long-wavelength gravitational
waves upon the polarization properties of  electromagnetic 
radiation as it propagates from cosmologically distant sources. 
Scalar (density) fluctuations do not rotate neither the plane 
of polarization of the electromagnetic radiation nor 
the source image. They produce, however, an appreciable 
shear, which distorts the image shape, leading 
to an apparent rotation of the image orientation
relative to its plane of polarization.
In sources with large ellipticity the apparent rotation 
is rather small, of the order (in radians) of the 
dimensionless shear. The effect is larger at smaller 
source eccentricity. A shear of 1\%  can induce 
apparent rotations of around 5 degrees in radio sources 
with the smallest eccentricity among those with a significant 
degree of integrated linear polarization.
We discuss the possibility that weak lensing by shear
with rms value around or below 5\% may be the cause for the 
dispersion in the direction of integrated linear polarization 
of cosmologically distant radio sources away from the 
perpendicular to their major axis, as expected from models 
for their magnetic fields. 
A rms shear larger than 5\% would be incompatible
with the observed correlation between polarization properties
and source orientation in distant radio galaxies and quasars.
Gravity waves do rotate both the plane of polarization as well 
as the source image. Their weak lensing effects, however,  
are negligible.

\end{abstract}


\section{Introduction}

Polarization properties of extragalactic radio sources are
primarily a  source of information about magnetic 
fields in the source, in the intergalactic medium, and in
the Milky Way (\cite{Saikia88,Kronberg94}). The orientation of 
the plane of linear polarization
at the source, derived from multi-frequency polarization 
measurements after correction for Faraday rotation 
along intervening magnetic fields, should point perpendicularly 
to the magnetic field lines, the emission being synchrotron radiation.

In the case of extended radio jets from quasars, the orientation
of the intrinsic linear polarization within a well defined
jet is typically perpendicular to the 
local direction of each jet segment, and if it
does change it is usually a $90\arcdeg$ flip
(see, e.g., \cite{Bridle86}; and \cite{Killeen86}).
Computational MHD modeling of radio jets 
improves the confidence in the predicted jet-intrinsic
magnetic field configurations supporting
these observations (\cite{Clarke89}).

Kronberg et al. (1991) have proposed and applied a technique involving
the angular relationship between the intrinsic polarization vectors and
the morphological structure of  extended radio jets  
to probe the gravitational distortion  of these sources by
foreground lensing galaxies. 
Since the polarization vector is
not rotated by the gravitational field 
of an ordinary gravitational lens, 
the bending of a jet away from its
intrinsic projected shape caused by the lensing effects
of intervening masses yields a detectable local ``alignment breaking''
between the polarization and the orientation of each jet segment.
Application of this idea to the analysis of the
radio and deep optical images of the strongly polarized
jet from 3C 9 provided the first 
estimate of 
the mass and mass distributions of two intervening galaxies
acting as lenses using this technique
(\cite{Kronberg91}; and \cite{Kronberg96}).

The notion that the plane of polarization
is not rotated by the gravitational
field of an ordinary gravitational lens, 
which underlies the method of Kronberg et al. (1991),
was addressed by Dyer \& Shaver (1992). Using symmetry
considerations they argued that the polarization
direction of photons in a beam under the influence
of a gravitational lens remains unaffected except
for lenses in very rapid rotation, an astrophysically
unlikely situation.
Similar conclusions were also reached by Faraoni (1993),
with a different approach.
The fact that the gravitational field of rotating bodies 
may rotate the plane of polarization of electromagnetic waves
has been addressed among others by Strotskii (1957),
Balazs (1958), Plebanski (1960), Mashhoon (1973), 
Pinneault \& Roeder (1977) and  Su \& Mallet (1980).

In the present work we analyse the effect of weak 
gravitational lensing
by large scale structure in the Universe upon the 
polarization properties
of electromagnetic radiation from distant sources. 
Through an extension of the method of 
``alignment breaking'', introduced by Kronberg et al.
(1991), we suggest that polarization properties
of distant radio sources may provide information
on the fluctuations in the gravitational
potential due to large scale density inhomogeneities.

When light from a distant source propagates across density 
inhomogeneities, the apparent shape and size of the source 
is distorted by tidal deflections. The effect is known as 
weak lensing, in contrast to the case when the bending of light rays
is so strong that multiple image formation is possible.
Weak lensing effects have already proved to
be a powerful tool 
to map the mass distribution in galactic halos  
(\cite{Valdes84}) and clusters of galaxies 
(\cite{Tyson90}). They are also a potential probe 
of density fluctuations on the largest scales 
(\cite{Blandford91,Miralda91,Kaiser92}), particularly
through the correlation in galactic ellipticities 
induced by the shear they produce. 
It has been estimated that
shear caused by large-scale density fluctuations in the linear 
regime is of the order of 1\%, depending on the 
cosmological model, the normalization of the power spectrum, and 
the redshift distribution of sources 
(\cite{Blandford91,Miralda91,Kaiser92}). 
Theoretical predictions on smaller scales (below $\sim$20 arcmin) 
by Jain \& Seljak (1996), show that 
density inhomogeneities in the non-linear regime 
induce a larger cosmic shear, of the order of a few percent. 
Recently, Schneider et al. (1997) presented the first 
evidence for the detection of a robust shear signal of about 3\% on 
scales of 1 arcmin, in agreement with the above theoretical 
expectations, from the analysis of data in a field containing the 
radio source PKS1508-05.
Long-wavelength gravitational waves (tensor metric perturbations)
are also a potential source of weak lensing of distant sources.

Radio sources with a significant degree of linear polarization,
are also usually elongated along one direction.
The angle $\chi-\psi$ between the intrinsic plane of 
polarization and the orientation of the structural axis has been
measured for a large number of sources, for which redshift information
is also available. The existence of a significant
correlation between the direction of integrated linear polarization 
and the source axis orientation has been noticed since the 
earliest investigations (\cite{Haves75,Clarke80}).  
For high redshift galaxies, a well defined peak 
is found around $\chi-\psi=90\arcdeg$, suggesting a magnetic
field parallel to the source axis.
For low redshift sources there is weaker correlation, 
with a narrow peak at $\chi-\psi = 0\arcdeg$, suggesting a magnetic 
field perpendicular to the source axis, and a broader peak at 
$\chi-\psi = 90\arcdeg$. It is believed that the $\chi-\psi 
= 90\arcdeg$ peak corresponds to high-luminosity sources, while 
the $\chi-\psi=0\arcdeg$ peak is ascribed to low-luminosity objects. 

The angle between the plane of integrated linear polarization
and the major axis of distant radio sources
is potentially a useful tool to search for unusual effects 
upon light propagation over cosmological distances, additional to 
the standard Faraday rotation. There have been, for instance, 
searches for correlations between the polarization
properties of extragalactic radio sources and their location,
which would evidence some kind of ``cosmological birefringence''.
While first attempts indicated that the residual polarization
rotation fitted a dipole rule 
(\cite{Birch82,Kendall84,Bietenholz84}), 
a later analysis of a  larger sample
by Bietenholz (1986) rejected that evidence. 
Recently, Nodland \& Ralston (1997) have claimed to find 
evidence in the data for a redshift-dependent dipole
anisotropy in the rotation of the plane of polarization. 
A more conventional interpretation and statistical
analysis of the same data (\cite{Carroll97,Eisenstein97}) 
rejects any statistically significant positive signal, 
which would also be in conflict with other potentially more 
accurate probes of such cosmological
rotation of polarization, through high resolution polarization
and intensity data of some specific radio sources 
(\cite{Leahy97,Wardle97}).

In this paper, we estimate the effect of weak lensing by
large scale structure, in the form of density fluctuations
or gravitational waves, upon the observed angle between the 
plane of polarization of the electromagnetic emission by distant 
radio sources and their image orientation.
We find it is in principle possible that the scattered
cases of distant radio sources which appear to be polarized
away from the direction either perpendicular or parallel to
the source axis, if Faraday rotation were the only propagation effect,
be actually a consequence of additional wavelength-independent
rotation of the polarization as their light travels to us
through an inhomogeneous universe.

In section 2 we review the formalism appropriate to describe
image distortions caused by scalar and tensor weak lensing. 
In section 3 we work in the geometric optics approximation  
and derive the rotation of the plane of 
polarization as light propagates in the presence of metric fluctuations. 
In section 4 we estimate the effect of large scale structure upon the 
angle $\chi-\psi$ between the plane of polarization and the orientation of
a source image, and show that a shear of 1\% can  change
the value of $\chi-\psi$ by $5\arcdeg$. 
In section 5, we discuss the possibility 
that scalar shear with a rms value 
around or below 5\%, which would cause apparent rotations
with a mean square dispersion of $20\arcdeg$, 
be the cause of the scattered cases of cosmologically 
distant radio sources with a direction of integrated linear polarization 
which is not perpendicular to their major axis. 
In section 6 we summarize our conclusions.

\section{Image distortion by random metric perturbations}

In this section we review the formalism appropriate to describe
weak lensing effects upon light propagation along cosmological
distances through a universe with large scale density
inhomogeneities and long wavelength gravitational waves. 
Part of the material in this section 
is not new (see for instance \cite{Kaiser92,Seljak94,Kaiser96a} and 
\cite{Kaiser96b}).
We include it for completeness, to define our method
and conventions, and because some of our results for the gravitational
distortions are at variance with those of Seljak (1994) and
Kaiser \& Jaffe (1997).
The differences do not conflict
with the main conclusion that weak lensing effects by large scale
density inhomogeneities accumulate over distances 
longer than the wavelength of the metric perturbation as
$(D/\lambda)^{3/2}$ (\cite{Seljak94}), while
weak lensing distortions by gravitational waves do not 
significantly accumulate (\cite{Kaiser96a}).
Even if the effect of gravitational waves is small,
it is relevant to precisely determine the focusing, shear and rotation
they induce if we want to determine the rotation of the
plane of polarization relative to the morphological structure of the
source.

We restrict our attention to effects that are linear in the
metric perturbations.  For simplicity, we shall consider scalar and
tensor metric fluctuations around a Minkowski background.
The generalization to a Robertson-Walker background is straightforward
(\cite{Kaiser96b}). The restriction to a Minkowski background 
oversimplifies the distance-dependence of the effects under 
consideration, but preserves their qualitative features as well as
the order of magnitude estimates. Vector perturbations could also
be easily incorporated into this formalism.

Consider a Minkowski background metric $\eta_{\mu \nu}$
with signature (-1,1,1,1), and linear perturbations $h_{\mu \nu}$. 
Greek indices
$\mu\nu,\alpha,\beta,... =0,1,2,3$
denote space-time coordinates. Latin indices
$i,j,...=1,2,3$ will
denote spatial coordinates,
but latin indices
$a,b,...=1,2$ will be reserved for the components
transverse to the unperturbed photons paths, that we shall
take as approximately parallel to the $x^3-$axis.
Boldface symbols will denote spatial 
vectors, i.e. {\bf m}, {\bf k},...

The linearized geodesic equations 
read,   
\begin{equation}{d^2{x}^{\mu}\over d z^2}=-\eta^{\mu\nu}
(h_{\nu\alpha ,\beta}-{1\over 2}
h_{\alpha\beta ,\nu}) p^{\alpha}p^{\beta}\quad . \label{xdd}
\end{equation}            
We denote the photons 4-momentum with $p^{\mu}=dx^{\mu}/dz$, 
where $z $ is an affine parameter (not necessarily coincident
with the $x^3$-coordinate).

We consider photon paths sufficiently close to the $x^3-$axis,
pointing towards an observer located at the origin of 
coordinates, and work not only up to linear order in
the metric perturbations, but also up to first order
in the photon departures away from the polar axis,
which we parametrize as 
\begin{equation}x^a(z)=\theta^a z +O(h,\theta^2)\quad\quad (a=1,2)\quad .
\end{equation}
We choose an affine parameter such that
$x^3=z+O(h,\theta^2)$, and thus $t=t_e + (z_s-z) +O(h,\theta^2)$,
with $t_e$ the time at which the photon is emitted at the source,
located at an affine distance $z_s$ in the $x^3$-direction.
We assume a gauge such that $h_{0j}=0$.
To lowest order $p^\mu =(-1,\theta^1,\theta^2,1)$, and
the geodesic equation for the transverse departure 
of the photon path away from the $x^3-$axis can be written as:
\begin{equation}
{d^2x^a\over dz^2}=D^a({\bf x},t)+{F^a}_b({\bf x},t)\theta^b
+O(\theta^2)
\end{equation}
where
\begin{equation}
\begin{array}{rl}
D_a &={1\over 2}(h_{00}+h_{33})_{,a}+h_{a3,0}-h_{a3,3}\\
F_{ab} &=h_{ab,0}-h_{ab,3}-(h_{3a,b}-h_{3b,a})\quad .
\end{array}
\end{equation}
Taylor-expanding $D_a({\bf x},t)$ to first order in $x^a$ 
we can finally write
\begin{equation}
{d^2x^a\over dz^2}=D^a(z)+{M^a}_b(z)\theta^b
+O(\theta^2)
\label{geo}
\end{equation}
where we defined
\begin{equation}
M_{ab}(z) =zD_{a,b}(z) +F_{ab}(z)\quad .
\label{Mab}
\end{equation}
The quantities in the r.h.s of equation (\ref{geo}) are 
evaluated over the $x^3$-axis and along the unperturbed photon path.
For instance
$D_a(z) =D_a(x^a=0,z,t=t_e+z_s-z)$.

In order to determine the distortion produced by metric
fluctuations upon light propagation, we consider
two neighboring photon paths
that arrive to the observer
with angular separation $\theta^a$
coming from two different points at the source separated by a
coordinate distance
$\Delta x^a(z_s)=\bar\theta^a z_s.$
The particular solution to the geodesic 
deviation equation with the appropriate boundary
conditions is
\begin{equation}
\Delta x^a(z)=\theta^a z+\int_0^{z}dz'\int_0^{z'} dz'' 
{M^a}_b(z'')\theta^b\quad .
\label{dxa}
\end{equation}
A source with a shape described in these coordinates
by $\Delta x^a=\bar\theta^a z_s$ is seen by an observer at the
origin as if the image were at $\Delta {\bar x}^a=\theta^a z_s$
in the source sphere.
The mapping
$\bar\theta^a\rightarrow\theta^a$ is given by
\begin{equation}
\theta^a=({\delta^a}_b+{\psi^a}_b)\bar\theta^b
\end{equation}
where
\begin{equation}
\psi_{ab}=
-\int_0^{z_s}dz\, {{z_s-z}\over z_s}\, M_{ab}(z)
\label{psi}
\end{equation}
is the distortion matrix, which describes the effect
of linear metric perturbations upon the observed shape of an 
image.

We point out now what is the source of variance between our results
and those by Seljak (1994) and Kaiser \& Jaffe (1997).
Their expressions 
for the geodesic deviation and 
$\psi_{ab}$ 
coincide with 
eqs. (\ref{dxa}) and 
(\ref{psi}) with $M_{ab}=zD_{a,b}$, rather than
the complete expression in eq. (\ref{Mab}).
The difference is the term $F_{ab}$, which arises
as a consequence that the photon paths
must converge into the origin, and thus the unperturbed paths
are not strictly parallel. The relevant observational situation
is not exactly described by the distortion suffered by
a bundle of initially parallel photon paths, but rather by
a bundle that starting at the source focalize at the observer
location. In other words, tidal deflections are not only due
to the gradients of the gravitational potential in the directions 
transverse to the central ray in the bundle, but there is 
also a contribution of the gradient in the longitudinal direction,
given that the rays in the bundle are not parallel. 
The longitudinal gradients, as we shall see, are significant in
the tensor case only, which at any rate is not likely to have
a large impact upon measurements.
The precise result for the focusing, shear 
and rotation induced by gravity waves, however, must take them into
account.
  
There are some subtleties in the definition of the distortion
matrix to linear order in the perturbations that we now discuss.
As defined by eq. ({\ref{psi}), the distortion matrix relates
coordinate (rather than proper) shapes. It is thus convenient to 
separate from $\psi_{ab}$ the local distortion effects, due to the 
different proper length that a same coordinate segment has at the 
observer and source locations. We thus define $\tilde\psi_{ab}$
such that
\begin{equation}
\psi_{ab}=\tilde\psi_{ab}+{1\over 2}\Delta h_{ab}\quad .
\label{psitilde}
\end{equation}
Here
$\Delta h_{ab}=h_{ab}(z_s,t_e)-h_{ab}(z=0,t=t_o)$
is the difference in the metric perturbations
between the events of emission of the light ray
at time $t_e$ from the source location, and of its 
observation at the origin  at time $t_o$.
$\tilde\psi_{ab}$ is the appropriate matrix
to describe the image distortion in terms of 
proper length mappings:
\begin{equation}
g_{ab}(0,t_o)\theta^a\theta^b\, z_s^2=[g_{ab}(z_s,t_e)+
\tilde{\psi}_{ab}]\ \bar\theta^a \bar\theta^b\, z_s^2\quad 
\end{equation}
with $g_{ab}=\eta_{ab}+h_{ab}$.

Although this is just a conventional
matter, we also wish to emphasize that our definition 
of focusing describes its dependence with affine,
rather than coordinate, distance. Since they differ
by terms linear in the perturbations, this matter of
convention needs to be specified.

The more general distortion matrix $\psi_{ab}$ can be decomposed 
in terms of a diagonal matrix, which represents focusing or 
convergence, a symmetric traceless matrix $\gamma_{ab}$, which 
describes shear, 
and an antisymmetric matrix, that corresponds to a rigid rotation
of the image:
\begin{equation}
\psi_{ab}=\kappa\,\delta_{ab} +\gamma_{ab} - \omega
\epsilon_{ab}\quad .
\label{psi-gral}
\end{equation}
Here $\epsilon_{ab}$ is the Levi-Civita tensor in two-dimensions.
Shear can be described in terms of its intensity $\gamma$ and 
the direction of one of its principal axis $\phi$ (which is only 
defined modulo $\pi$) as
\begin{equation}
\gamma_{ab}=\gamma\left ( \begin{array}{cc}
\cos 2\phi & \ \sin 2\phi\\
\sin 2\phi & -\cos 2\phi
\end{array}\right )\quad .
\label{gamma}
\end{equation}
Notice that a rotation of the direction of the shear 
by $\pi/2$ amounts to just a change in the sign of $\gamma$. 
Two independent directions in which to decompose an arbitrary 
shear form an angle $\pi/4$.

We now specialize the expressions above to the case of
scalar and tensor metric perturbations respectively.
We start working with a single planar metric 
fluctuation, and later compute the rms effect of a
full stochastic background by superposition.

\subsection{Scalar perturbations}

We consider one Fourier mode of static scalar perturbations
\begin{equation}
h_{\alpha\beta}({\bf x})
=h({\bf k}) e^{i{\bf k\cdot x}}\delta_{\alpha\beta}
=h({\bf k})e^{ik_ax^a}e^{-ik\mu z}\delta_{\alpha\beta}
\end{equation}
appropriate for instance to describe adiabatic density 
perturbations in a matter-dominated, spatially-flat 
Robertson-Walker background, in the longitudinal gauge.  
We have defined 
$\mu={\bf m\cdot n}$,
where $\mu$ is the cosine of
the angle between the wave-vector ${\bf k}=k{\bf m}$,
and the direction ${\bf n}$ in which light propagates,
which is approximately along the $x^3-$axis, incoming 
towards the origin from positive values of $x^3$.
Specialization of the eqs. in the previous section to this case
lead to
\begin{equation}
\psi_{ab}=h({\bf k})[\alpha(\mu,z_s)m_am_b+\beta(\mu,z_s)
\mu\delta_{ab}]\quad .
\label{psi-esc}
\end{equation}
Here we defined
\begin{equation}
\begin{array}{rl}
\alpha(\mu,z_s) &=
{k^2}\int_0^{z_s}dz {{z(z_s-z)}\over z_s}e^{-ik\mu z} \\
\beta(\mu,z_s) &=
{-ik}\int_0^{z_s}dz {(z_s-z)\over z_s} e^{-ik\mu z}
\quad .
\end{array}
\label{abe}
\end{equation}
We write the projection of the
wave-vector direction onto the plane orthogonal to $z$ as 
$m^a=\sqrt{1-\mu^2}(\cos\varphi,\sin\varphi)$,
and from eqs. (\ref{psi-gral}) and (\ref{psi-esc}) we
find that the focusing, shear intensity and direction,
and rotation induced by a single Fourier mode of scalar perturbations
are
\begin{equation}
\begin{array}{rl}
\kappa &={1\over 2}h({\bf k})[(1-\mu^2)\alpha(\mu,z_s)+2\mu
\beta(\mu,z_s)]\\
\gamma &={1\over 2}h({\bf k})\ (1-\mu^2)\alpha(\mu,z_s)\\
\phi &=\varphi\\
\omega &=0 \quad .
\end{array}
\label{psis}
\end{equation}
Scalar perturbations induce no rotation. 
They produce shear in 
the direction of the perturbation wave-vector, 
projected into the source plane.

As pointed out by Kaiser \& Jaffe (1997), the reason why 
weak lensing
effects by scalar perturbations accumulate over distances
longer than the perturbation wavelength is evident from eqs.
(\ref{abe}) and (\ref{psis}).
Resonant modes, those for which the photons
remain in phase with the perturbation, are those with 
$\mu\approx 0$, which produce a significant deflection.
Notice that $\alpha(\mu)\rightarrow (kz_s)^2/6$ and 
$\beta (\mu)\rightarrow ikz_s/2$ as $\mu\rightarrow 0$.
Our results differ from those by 
Seljak (1994) and 
Kaiser \& Jaffe (1997) by the 
effect
of the longitudinal modes 
which affect just the focusing $\kappa$
by the term proportional to
$\mu\beta(\mu,z_s)$. Since this 
term does not contribute in the resonant situation ($\mu=0$),
it is sub-dominant, and negligible, after superposition
of all Fourier modes of the stochastic background.

As discussed in the previous subsection,
the local effects can be subtracted
from the distortion matrix describing proper
distortions through 
$\tilde\psi_{ab}=\psi_{ab}-{1\over 2}\Delta h_{ab}$. 
In this case
$\Delta h_{ab}=\delta_{ab}(e^{-ik\mu z_s}-1)$,
and use of the relation $2\mu\beta-\mu^2\alpha=
(e^{-ik\mu z_s}-1)$ allows to identify the
proper distortion matrix elements as  
\begin{equation}
\begin{array}{rl}
\tilde\kappa&={1\over 2}h({\bf k})\alpha(\mu,z_s)\cr
\tilde\gamma&=\gamma \quad .
\end{array}
\end{equation}

Consider now a stochastic background of scalar 
fluctuations 
characterized 
in momentum-space by the correlations 
$<h({\bf k})h^*({\bf k}')>=
(2\pi)^3 \delta({\bf k}-{\bf k}')P(k)$
where $P(k)$ is the power spectrum.
The mean square expectation value of the shear induced is
\begin{equation}
<\gamma^2>=\int {dk\over (2\pi)^3}k^2 P(k)
{\pi\over 2}\int_{-1}^1 d\mu  
\ (1-\mu^2)^2\vert\alpha\vert^2 
\label{gammae}
\end{equation}
In the limit $kz_s\gg  1$ the $\mu$-integral equals
${\pi\over 15}\ (kz_s)^3 +O[(kz_s)^{2}]$. Shear induced by 
scalar perturbations of amplitude $h$ accumulates over distances
$D$ longer than the wavelength $\lambda$ as $h\times
(D/\lambda )^{3/2}$ (Seljak 1994).
In the case of a scale invariant power spectrum $P(k)$,
such as that predicted by inflationary models in a CDM scenario,
with amplitude, normalized at small $k$
according to the COBE-DMR measurement of the cosmic microwave 
background anisotropy, the rms shear is of order 
$<\gamma^2>^{1/2}\approx 0.02$. The rms convergence is of the same
order of magnitude.

\subsection{Tensor perturbations}

Now we study image distortions produced by gravity waves.
We work tensor perturbations in the transverse traceless gauge,
where $h_{\alpha\beta} k^\alpha =0$, $h_{0\mu}=0$, and $h^\mu_\mu 
=0$. We consider gravitational waves with wave vector 
$k^\mu=k(1,{\bf m})$, with  $m^i=(\sqrt{1-\mu^2}\cos\varphi,
\sqrt{1-\mu^2}\sin\varphi,-\mu)$. As in the previous section,
$\mu={\bf m\cdot n}$ is the cosine of the angle between 
the perturbation wave-vector and the photon propagation direction.
We denote  the two independent polarization modes of the gravity
waves with the symbols $+,\times$. The normal modes of the
metric perturbations read 
$h_{ij}({\bf x},t)=h_{ij}({\bf k})e^{i({\bf k\cdot x} - kt)}$
with
\begin{equation}\begin{array}{rl}
h_{11}({\bf k}) &= h_+({\bf k})(\mu^2\cos^2\varphi-\sin^2\varphi) 
-h_\times({\bf k})\mu\sin2\varphi\\
h_{22}({\bf k}) &= h_+({\bf k})(\mu^2\sin^2\varphi-\cos^2\varphi) +
h_\times({\bf k})\mu\sin2\varphi\\
h_{33}({\bf k}) &= h_+({\bf k})(1-\mu^2)\\
h_{12}({\bf k}) &=h_+({\bf k}){1\over 2}(1+\mu^2)\sin2\varphi
+h_\times({\bf k})\mu\cos2\varphi\\
h_{13}({\bf k}) &=h_+({\bf k})\mu\sqrt{1-\mu^2}\cos\varphi
-h_\times({\bf k})\sqrt{1-\mu^2}\sin\varphi\\
h_{23}({\bf k}) &=h_+({\bf k})\mu\sqrt{1-\mu^2}\sin\varphi
+h_\times({\bf k})\sqrt{1-\mu^2}\cos\varphi \quad .
\end{array}
\end{equation}
The distortion matrix now reads, from eq. (\ref{psi})
\begin{equation}\begin{array}{rl}
\psi_{ab} =&
\alpha(\mu-1,z_s)\left [ {1\over 2}h_{33}m_am_b - (1-\mu)
h_{3a}m_b \right ] - \\
& \beta(\mu-1,z_s)\left [(1-\mu)h_{ab} +(h_{3a}m_b-h_{3b}m_a) 
\right ]
\end{array}
\end{equation}
where $\alpha$ and $\beta$ are exactly the same functions as in
eq. (\ref{abe}), now the argument being  $\mu-1$ instead of 
$\mu$. We have discarded an irrelevant overall phase factor.
The focusing, shear intensity and direction, and rotation
induced by each normal mode of a gravity wave background,
for each independent polarization, thus turn out to be
\begin{equation}\begin{array}{rl}
\kappa\ &= {1\over 4}(1-\mu^2)h_+({\bf k})\left [ (1-\mu)^2
\alpha(\mu-1,z_s)+2(1-\mu)\beta(\mu-1,z_s)\right ] \\
\gamma_+ &= {1\over 4}h_+({\bf k})\left [ (1-\mu^2)(1-\mu)^2
\alpha(\mu-1,z_s)-2(1+\mu^2)(1-\mu)\beta(\mu-1,z_s)\right ]\\
\phi_+ &=\varphi\\
\gamma_\times &= {1\over 2}h_\times({\bf k})\left [ 
(1-\mu^2)(1-\mu)
\alpha(\mu-1,z_s)+2\mu(1-\mu)\beta(\mu-1,z_s)\right ]\\
\phi_\times &=\varphi-{\pi\over 4}\\
\omega\ &=-{1\over 2}(1-\mu^2)h_\times({\bf k})\left [
(1-\mu)\alpha(\mu-1,z_s)+2\beta(\mu-1,z_s)\right ] \quad .
\end{array}
\label{psit}
\end{equation}

Gravity waves have quite different weak lensing effects than
scalar metric perturbations. The fact that waves propagate 
imply that the resonant modes, those that may eventually add-up 
over distances longer than the wavelength, are now those 
with $\mu\approx 1$, which do not produce neither focusing, 
shear, nor rotation. Eqs. (\ref{psit})
coincide with the equivalent expressions of Kaiser \&
Jaffe (1997), except for the terms proportional to 
$\beta(\mu-1,z_s)$.

The $+$ polarization of the gravity waves produces focusing,
shear with a principal axis in the direction of 
the wave-vector projection into the source plane ($\varphi$), 
and no rotation.
The $\times$ polarization produces no focusing, a shear comparable 
in strength to that of the $+$ polarization but  with 
principal axis at $45\arcdeg$, and induces rigid rotation of
the image. In other words, the $+$ polarization produces similar
effects as scalar shear, while the $\times$ polarization induces
truly pseudo-scalar shear (the change from a right-handed 
coordinate system to a left-handed one changes the sign of
$h_\times$ but not of $h_+$).

We now use the relation
$2(1-\mu)\beta+(1-\mu)^2\alpha=1-e^{ikz_s(1-\mu)}$
to express the rotation $\omega$ as 
\begin{equation}\omega = {1\over 2} (1+\mu)\Delta h_\times({\bf k})
\end{equation}
where
$\Delta h_\times({\bf k})=h_\times({\bf k})(e^{ik(1-\mu)z_s}-1)$
is the variation in the amplitude of the $\times$ component of
the gravitational wave between the events of emission of the light
ray at the source and its arrival to the observer at the origin.
Clearly, there is no possible cumulative effect. The rotation
can never exceed the dimensionless wave amplitude.

To separate the local distortion effects from those
produced during light propagation we evaluate
$\tilde\psi_{ab}=\psi_{ab}-{1\over 2}\Delta h_{ab}$,
in this case the proper distortion matrix elements read:
\begin{equation}\begin{array}{rl}
\tilde\kappa &=0\cr
\tilde\gamma_{(+,\times)} &={1\over 2}(1-\mu)^2\alpha
(\mu - 1,z_s)h_{(+,\times)}\cr 
\tilde\omega&=\omega \quad .
\end{array}
\label{gammat}
\end{equation}
Expressed in terms of proper areas and as a function of
affine distance to the source, there is no focusing to
linear order in the gravity wave amplitude, in agreement
with previous conclusions by Zipoy \& Bertotti (1968).

Consider now a stochastic background of gravitational 
radiation, built as a superposition of plane sinusoidal waves.
Contrary to the scalar case, they  induce distortions just of 
the order of their dimensionless amplitude, but not larger
(Kaiser \& Jaffe 1997).
Consider, for instance, an stochastic background such as that
predicted by inflationary cosmological models (\cite{Abbott84}),
(rms amplitude proportional to the wavelength),
with the largest possible amplitude compatible with the observed
CMB anisotropy. The rms amplitude is of the order of 
$h(k)\sim 10^{-6}H_o/k$. The dominant effect is now that of
the wavelengths comparable to the distance to the source.
Distortions of order $\gamma\simeq 10^{-6}$, a
factor $10^{-4}$ smaller than the effect expected to be induced 
by scalar perturbations, are unlikely to be detected.
At any rate, it is interesting to observe that there are
specific footprints of weak lensing by tensor perturbations,
due to its pseudo-scalar contribution, which may eventually serve 
to distinguish it from that produced by density inhomogeneities.

\section{Rotation of the plane of polarization by gravity waves}

One specific footprint of weak lensing by gravity waves is
rotation, which is not induced by scalar perturbations.
Rotation of an image may be difficult to ascertain, without
a knowledge of the intrinsic source orientation. The polarization
properties of the electromagnetic emission of the source may, in
some cases, bear some correlation with the intrinsic source 
orientation. The angle between some structural direction in
a source and the direction of the plane of polarization is thus
potentially a useful tool to search for eventual image rotation by
weak lensing. Since the plane of polarization may also rotate
as electromagnetic radiation propagates across density inhomogeneities
or gravitational waves, we should also estimate this effect.
To do so, in this section we work within the geometric optics
approximation of the solution to Maxwell equations in curved space-time 
(\cite{MTW}), 
which imply that the direction of linear polarization is 
parallel-transported along the photon path. 
It is intuitive that scalar perturbations, which have no handedness, 
do not rotate the plane of polarization. In the case of gravity waves, 
however, we may
expect them to induce a ``gravitational Faraday effect'' 
(\cite{Perlick93,Cooperstock93}). 
Image rotation being a consequence of
the geodesic deviation between neighboring rays while 
polarization rotation arises from parallel transport, we may
in principle expect them to be different, and thus potentially 
provide a test to detect absolute rotations.

Typical wavelenghs of electromagnetic radiation 
are extremely short compared to the radius of 
curvature of space-time and to the typical
length over which the amplitude, polarization, and wavelength 
of the electromagnetic waves may vary. Thus, we may
confidently apply the geometric optics approximation in our analysis, 
and electromagnetic radiation 
may be regarded locally as plane waves that propagate through
a space-time of negligible curvature.
In the Lorentz gauge the electromagnetic vector potential may 
be written as  
\begin{equation}
A^\mu=\Re{\{ a^\mu e^{i\theta}\}}
\label{a}
\end{equation}
where $a^\mu$ is a vector amplitude and $\theta $ is a real 
phase that rapidly changes, with the typical frequency of 
the electromagnetic radiation, $\theta=\omega p^\alpha x_\alpha 
+ {\rm const.}$, with $\omega$ the 
wave frequency and $p^\alpha$ the dimensionless wave-vector.
Expanding the solution of Maxwell's equations in powers
of the electromagnetic wavelength divided by the length-scale
over which the gravitational potentials vary significantly
(the gravitational wave wavelength in our case)
one finds to lowest order that the photon paths are
null geodesic, and to the next order that
the vector amplitude $a^\mu$ can be written as 
$a^\mu =af^\mu$ where
$a$ is a scalar amplitude and $f^\mu$ is the 
polarization vector (normalized as $f^\mu f_\mu=1$), 
which satisfies the equation of parallel transport 
along the photon path. 

The electric field $E^j$ measured by a comoving observer 
is derived from the
electromagnetic field tensor $F^{\mu\nu}=A^{\nu ;\mu}-A^{\mu;\nu}$
as 
\begin{equation}
E^j=F^{0j}=\Re{\{ [(a^{j,0}-a^{0,j})+
i\omega (a^jp^0-a^0p^j)]e^{i\theta}\}}
\quad .
\label{ejo}
\end{equation}
The length-scale over which the vector amplitude varies being much
longer than the electromagnetic wavelength, 
the first term in the r.h.s. of Eq.(\ref{ejo}) is negligible 
compared to the second. Thus, the direction of the electric
field is indeed given by the direction of $a^i$, in turn
determined by $f^i$.

We shall calculate the projection of the electric field direction into
the plane perpendicular to the $x^3$-axis, $f^a$ with 
$a=1,2$.  
Even though the photon paths 
deviate away from the polar axis in the presence of metric fluctuations,
with typical deflections of order $h$,
the projection into the plane perpendicular to unperturbed
photon paths differs from the projection into the plane
transverse to the actual photon trajectory 
by terms of order $h^2$.    

The transverse components of the polarization vector
verify the parallel-transport equation, which along a ray
that propagates nearly parallel to the $x^3-$axis towards the
origin read:
\begin{equation}
{df^a\over dz}={1\over 2}\delta^{ab}(h_{bc,0}-h_{bc,3} +
 h_{3c,b}-h_{3b,c})f^c\quad .
\label{pt}
\end{equation}
Notice that, to the order considered, the difference
between the affine parameter $z$ and the $x^3-$coordinate
is now irrelevant.

In the case of scalar static perturbations, $h_{ab,3}=h_{,3}
\delta_{ab}$ is the only non-zero term in the r.h.s. of this
equation. The solution to eq. (\ref{pt}) is such that
\begin{equation}
f^a(z=0)=[{\delta^a}_b +{1\over 2}\Delta{h^a}_b] f^b(z_s)\quad .
\end{equation}
Since $\Delta h_{ab}$ is diagonal, $f^1$ and
$f^2$ are modified at the same rate, and thus there is
no rotation of the plane of
polarization with respect to the fixed coordinate grid.
In other words: if 
\begin{equation}
\chi(z)=\arctan \left ({f^2(z)\over f^1(z)}\right )
\end{equation}
and $\Delta\chi=\chi(z_s)-\chi(0)$, then 
\begin{equation}
\Delta\chi=0\quad {\rm (scalar)}\quad .
\label{deltachis}
\end{equation}

In the presence of a plane and monochromatic 
gravitational wave with wave-vector ${\bf k}=k{\bf m}$,  eq.
(\ref{pt}) has instead non-diagonal terms
\begin{equation}
{df^a\over dz}={-ik\over 2}e^{ikz(1-\mu)}
[{h_3}^a({\bf k})m_b -h_{3b}({\bf k})m^a
+ (1-\mu){h^a}_b({\bf k})]f^b\quad .
\end{equation}
Integration of this equation between the source and the
observer leads to
\begin{equation}
f^a(z=0)=[{\delta^a}_b  -\omega{\epsilon^a}_b
+ {1\over 2}\Delta {h^a}_b] f^b(z_s)
\label{ft}
\end{equation}
where, as in the previous section, $\Delta h_{ab}=
h_{ab}(z_s,t_e)-h_{ab}(0,t_o)$ denotes the variation 
in the gravitational wave amplitude between the events 
of emission and observation of the photons,
and $\omega={1\over 2}(1+\mu)\Delta h_\times$
coincides with the value for the image rotation
induced by gravity waves. The angle that the polarization
forms with respect to the $x^1$ -axis in this fixed
coordinate grid thus changes by the amount
\begin{equation}
\Delta\chi=\omega   +
{1\over 4}(1+\mu^2)\Delta h_+\sin 2(\varphi-\chi)
+{1\over 2}\mu \Delta h_\times\cos 2(\varphi-\chi) 
\quad {\rm (tensor)}\quad .
\label{deltachit}
\end{equation}
The first term in the r.h.s, independent of
the initial direction of polarization, 
coincides with the image rotation $\omega$
derived in the previous section. 
The other two terms, that arise from the presence of
$\Delta h_{ab}$ in eq. (\ref{ft}),
coincide with the apparent rotation of a direction that 
subtends an angle $\chi$ with respect to  the $x^1$-axis 
as a consequence of the local variation in the gravitational 
potentials between the source and observer locations
(as we shall discuss further in the next section). 

Clearly,  the rotation of the plane of
polarization does not accumulate over distances longer than the
wavelengths as light travels through a stochastic background 
of gravitational waves, since the effect is proportional to
the change in the wave amplitude between the emission and
observation events. 
The  rotation of the plane of polarization produced by a stochastic
background of gravitational waves of cosmological wavelengths and
amplitude of order $10^{-6}$ is thus totally negligible.

\section{\bf Apparent rotations}

The rotation  of the plane of polarization as light
propagates across gravity waves derived in
the previous section, or the lack of it in the case
of scalar perturbations, were measured with respect to
the fixed coordinate grid $(x^1,x^2,x^3)$. 
Notice, however, that due to weak lensing shear and rotation, 
be it of scalar or tensor nature,
light originated in a point located at
$\bar\theta^a z_s$, in a direction that forms an angle
$\xi=\arctan(\bar\theta^2/\bar\theta^1)$
with respect to the $x^1$-axis in the fixed coordinate grid,
is seen by the observer at the origin 
as coming from a direction, in this same
coordinate grid, at an angle 
$\xi+\Delta\xi=\arctan(\theta^2/
\theta^1)$, with
\begin{equation}
\Delta\xi=\omega+\gamma\sin2(\phi - \xi)\quad .
\label{dxi}
\end{equation}
We recall that $\gamma$ is the shear intensity
and $\phi$ its direction.
A given fixed direction at the source is
seen by the observer not just
rotated by $\omega$ with respect to the absolute direction
in this Euclidean grid, but is also affected by shear.
In the case of scalar fluctuations, the apparent
rotation of a fixed direction $\xi$  induced by a
single Fourier mode is
\begin{equation}
\Delta\xi=\tilde\gamma\sin2(\varphi - \xi)\quad {\rm (scalar)}.
\label{dxiscalar}
\end{equation}
In the case of gravity waves, it is given by
\begin{equation}
\begin{array}{rcl}
\Delta\xi=&\omega
+ {1\over 4}(1+\mu^2)\Delta h_+\sin 2(\varphi-\xi)
+{1\over 2}\mu \Delta h_\times\cos 2(\varphi-\xi) & \quad \cr
&+ \tilde\gamma_+\sin2(\varphi - \xi)
-\tilde\gamma_\times\cos2(\varphi-\xi)
& \quad {\rm (tensor)}.
\end{array}
\label{dxitensor}
\end{equation}
The first term in the r.h.s. of this equation 
is the rigid rotation $\omega$. The next two terms 
are the effect upon a given fixed coordinate
direction of the change in the gravitational
potential between the emission and observation 
events. The rest is the additional apparent rotation
due to the proper shear $\tilde\gamma$, given by eq.
(\ref{gammat}).

The question to address is what are the potentially
observable consequences of apparent rotations of 
directions in the source relative to its 
plane of polarization.
A possible approach is to consider sources 
with some structural, intrinsic direction. We can then answer 
the question of what happens to the apparent angle between the
plane of polarization and that structural direction as
light from the source propagates across metric perturbations.  

Consider sources  with a significant degree of linear
polarization and elongated shape,
with approximately elliptical isophotes.
For these objects we denote with $\psi$ the position angle
on the sky  (the orientation of the major axis of the image),
and with $\chi$ the direction of the plane of
integrated linear polarization, after subtraction of
the wavelength-dependent Faraday rotation along intervening
magnetic fields.

Weak lensing by metric perturbations
produces focusing, shear, and rotation. 
Convergence amplifies  sources
without change in their shape, so it does not affect 
the observed position angle $\psi$.
Rotation, as described by $\omega$ in
eq. (\ref{psi-gral}),
accounts for a rigid rotation of the image, with the 
distance between any pair of points unchanged. 
Tensor perturbations rigidly rotate 
the position angle $\psi$ by the amount $\omega$. 
However, even when no rigid
rotation occurs, as is the case for scalar perturbations, 
shear distorts a source image in such a way that
an ellipse is still seen as an ellipse but with different 
ellipticity and orientation.

Consider an elliptical source 
whose principal axis forms an angle $\psi$ with respect to
the $x^1$-axis in the source plane.
$\bar{x}^a=\bar\theta^az_s$ is taken to describe such an ellipse.
We denote the length of the major and minor axes by 
$a$ and $b$ respectively. 
We define the ellipticity of the source as 
$\epsilon =(a^2-b^2)/(a^2+b^2)$.
When light from this object arrives to the observer
distorted by convergence, shear, and rotation,
the image parametrization ${x}^a=({\delta^a}_b + 
{\psi^a}_b)\bar\theta^b z_s$ in the observer sphere 
still describes an ellipse, but rotated an angle 
$\Delta\psi$ and with a different ellipticity 
$\epsilon '$, given by (if the distortion is small):
\begin{equation}
\tan (2\Delta\psi)={2[\omega +(\gamma/\epsilon)\sin 2(\phi-\psi)]\over 
{1+ 2(\gamma/\epsilon)\cos 2(\phi-\psi)}}
\label{deltaxi}
\end{equation} 
\begin{equation}
\epsilon '=
{\sqrt{[\epsilon+2\gamma\cos 2(\phi-\psi) ]^2+4[\epsilon \omega +
\gamma\sin 2(\phi-\psi) ]^2}
\over {1+2\epsilon \gamma\cos 2(\phi-\psi) }}\quad .
\label{epsilon'}  
\end{equation}
Here $\phi$ is, as in previous sections, the direction of the
shear principal axis. If the ellipticity is larger than the 
shear, $\epsilon\gg\gamma$, we can approximate
\begin{equation}
\Delta\psi\approx \omega+{\gamma\over\epsilon}\sin 2(\phi-\psi)
\label{deltaxis}
\end{equation}
(which of course coincides with eq. (\ref{dxi}) in the
limit $\epsilon=1$),
and the new ellipticity is 
\begin{equation}\epsilon'\approx \epsilon+ 2\gamma (1-\epsilon)\cos 
2(\phi - \psi)\quad .
\end{equation}
In the opposite limit, of negligible eccentricity,
$\epsilon'\approx 2\gamma$ and $\Delta\psi\approx\phi$,
which simply means that the nearly circular source is distorted
in the direction of the shear.

We conclude that the angle between the plane of integrated
linear polarization and the source axis, $\chi-\psi$, 
is observed 
to change, as light propagates across density inhomogeneities
and gravitational waves, by the difference $\Delta(\chi -\psi)$
as given by
eqs. (\ref{deltachis},\ref{deltachit}) and (\ref{deltaxi}). 

Rotation and shear produced by a cosmological
background of gravitational waves, at most
of order $10^{-6}$, lead to a
totally negligible apparent rotation between the
plane of polarization and the source position angle.

Scalar shear leads to an apparent rotation
$\Delta(\chi-\psi)=-\Delta\psi$, since $\Delta\chi=0$ for
scalar perturbations.
In sources  with large ellipticity ($\epsilon \approx 1$),
the apparent rotation is of the order of the shear itself.
Such a small apparent rotation, at most of the order of a
few degrees, is not likely to have significant observational 
consequences.
Gravitational distortions due to weak lensing by large-scale structure
will produce no observable effect on the polarization properties 
of extended radio jets in quasars.
In the opposite limit, nearly circular sources appear 
distorted in the direction of the scalar shear,
while their polarization plane remains in the same direction.
Scalar shear is thus able to mask any correlation
between  orientation and polarization that a nearly
circular source might have. 

The most interesting regime may be the case of
sources with ellipticities of the order of
$\epsilon\approx 0.1$. 
These are the smallest ellipticities among radio sources
that still present some significant (larger than 5\%)
degree of linear polarization. They
can display an apparent rotation of the plane of
polarization relative to their major axis of order
\begin{equation}\Delta(\chi-\psi) \approx {\gamma\over\epsilon}
\quad .
\label{concl}
\end{equation}
A scalar shear of 1\%  can
change the observed angle between polarization and source
orientation by 5 degrees.

\section{Effects of shear upon the correlation between 
source polarization and orientation} 

The  direction of the plane of 
polarization in radio galaxies and quasars 
is expected to trace the geometric configuration 
of the magnetic field in such objects, at least in sources
with a significant degree of linear polarization. 
Discrepancies between the observed values 
and expectations based in models for the magnetic field
in distant radio sources could eventually signal the
presence of unusual effects additional to the standard Faraday
rotation in the polarization properties of light 
as it propagates through the universe
(\cite{Birch82,Carroll90,Harari92,Nodland97,Carroll97}).

It is possible that weak lensing by scalar perturbations 
explain the dispersion in the direction of integrated
polarization of  cosmologically distant 
radio sources away from the perpendicular to their 
major axis. We now discuss this possibility.

Consider the data on 160 radio sources used in
Carroll et. al (1990), with the corrections noted
in Carroll and Field (1997).
There are 89 sources with redshift $z<0.3$ and
71 with redshift $z>0.3$.  
The histogram of number of radio sources vs. $\chi - \psi$ 
in the case of nearby ($z<0.3$) galaxies does not 
display a very strong correlation, although there is a 
narrow enhancement around $\chi-\psi\approx 0\arcdeg$ and
a broader peak at $\chi-\psi\approx 90\arcdeg$.
In the case of distant ($z>0.3$) sources, the peak 
around $\chi-\psi\approx 90\arcdeg$ enhances dramatically, 
while there is no noticeabe peak around $\chi-\psi\approx 
0\arcdeg$. It is likely that any correlation between 
polarization and source orientation be weaker in the case
of sources only weakly polarized. Restriction of the
sample to the case of sources with maximum polarization 
$p_{\rm max}>5\%$ does indeed significantly
enhance the two peaks in the histogram of nearby sources,
while the peak around $\chi-\psi\approx 90\arcdeg$ 
for distant sources becomes even more convincing.

The most likely explanation for these observations
is that the sample consists of two different populations.
One population would consist of low luminosity sources, 
for which either there is no significant correlation
between polarization and position angle, particularly in
the case of weakly polarized sources, or have 
$\chi-\psi\approx 0\arcdeg$.
The low luminosity population would be underepresented in
the high redshift observations, where the second class,
of high luminosity sources with polarization almost 
perpendicular to their axis, would dominate 
(\cite{Carroll97,Clarke80}).

We now put forward the following scenario, which appears
to be consistent with present observations. Assume that
most of the distant radio sources that have a significant
degree of integrated linear polarization are polarized
in a direction almost perpendicular to their major axis
(say less than $5\arcdeg$ away from it), while a small
fraction does not posses a significant correlation 
between polarization and position angle. Scalar shear 
then acts stochastically upon the light emitted by these 
sources. The result should be an increase in the spread
of the observed $\chi-\psi$ away from $90\arcdeg$,
proportional to the rms value of the shear. The effect
actually depends upon the source redshift and ellipticity.
If the number of data points were sufficiently
large one could search for a correlation between
departures of $\chi-\psi$ away from $90\arcdeg$ with
the source redshift and ellipticity, and put to test
this scenario. Present observations are not sufficient 
to make serious attempts in this direction, but we 
shall argue that they are at least consistent with 
the above scenario, if the scalar shear has rms value 
around 5\%. Notice that this shear needs to be coherent 
only over small scales, of the order of the angular 
size of the source.

Given that the number of data points is not sufficiently
large to search for any correlation of the effect of shear
with the source redshift and ellipticity,
we shall estimate the effect to be expected in a 
population of radio sources with ellipticity equal
to the mean in the sample, at a redshift equivalent
to the mean redshift. Since shear effects do not 
significantly accumulate in the case of nearby sources,
we choose to consider just the 51 sources with $z>0.3$ 
and maximum integrated linear polarization 
$p_{\rm max}> 5\%$. 
We plot the corresponding histogram in Figure 1.
These sources have a mean value 
of ${\chi-\psi}$ equal to $90\arcdeg$ with rms dispersion 
$\sigma =33\arcdeg$, and mean redshift $\bar z=0.85$.
Of these 51 sources, 41 have their angle of 
intrinsic polarization less than $45\arcdeg$ away from the direction
perpendicular to the source axis, while the remaining 10
appear to be more or less uniformly distributed at angles
less than $45\arcdeg$ away from the source axis,
although the number is too small to draw significant
conclusions. If we exclude from the sample these 10 sources,
under the assumption that they may be predominantly members
of a population with weak correlation between polarization
and position angles, the remaining 41 sources have a mean
value of ${\chi-\psi}$ of $91\arcdeg$ with rms dispersion
$\sigma=19\arcdeg$ and mean redshift $\bar z=0.84$.  
We have evaluated
the ellipticites of all but 4 of them, for which we
found the values of the major and minor axis 
quoted in Gregory at. al (1996). The mean ellipticity 
of the sample is $\bar\epsilon\approx 0.15$. 
We put forward the hypothesis that the intrinsic 
dispersion of $\chi-\psi$ away from $90\arcdeg$ is 
much smaller than the observed value, the observed
dispersion being the consequence of the
action  of stochastic cosmic shear. 

We now estimate the amount of shear necessary to 
produce a mean square rotation $\sigma\approx 20\arcdeg$
upon a population of sources with ellipticity 
$\epsilon=0.15$, assumed to be located at equal redshift.
In the limit $\bar\gamma/\epsilon <<1$, where 
$\bar\gamma$ is the rms value of the shear, assumed to
act in all directions in the plane perpendicular to the 
photon paths with uniform probability, the results of
the previous section (eq. (\ref{deltaxis})) indicate 
that the rms value of the apparent rotation is
\begin{equation}
\sigma_\psi\equiv <(\Delta\psi)^2>^{1/2}
\approx \left [ {2\over \pi}\int_0^{\pi/2}d\phi
\left ({\bar\gamma\over\epsilon}\right )^2\sin^2 2\phi
\right ]^{1/2}={1\over\sqrt 2}{\bar\gamma\over\epsilon}
\quad .
\label{rmssmall}
\end{equation} 
A value $\sigma_\psi=0.35$ ($20\arcdeg$) 
corresponds to $\bar\gamma/\epsilon\approx 0.5$,
which implies a rms shear of 7.5\% if the  
ellipticity of the sources is $\epsilon=0.15$.  
Eq. (\ref{deltaxis}), however, underestimates the
apparent rotation when  $\gamma/\epsilon$
is not significantly smaller than unity,
as is frequently the case if the rms value 
is $\bar\gamma /\epsilon=0.5$. 
To better estimate the rms apparent 
rotation in a general case we should use eq. 
(\ref{deltaxi}). Assuming a gaussian distribution
of $\gamma$ around a zero mean, with rms
dispersion $\bar\gamma$, and uniform probability 
distribution in the shear direction $\phi$, the
rms rotation reads
\begin{equation}
\sigma_\psi=
\left [ {2\over \pi}\int_0^{\pi/2}d\phi
\int_{-\infty}^{+\infty}d\gamma
{1\over 4}\arctan^2 \left ( {2(\gamma/\epsilon)
\sin 2\phi\over 1+2(\gamma/\epsilon)\cos 2\phi}\right )
{e^{-\gamma^2/2{\bar\gamma}^2}\over \sqrt{2\pi}\bar\gamma}
\right ]^{1/2} 
\quad .
\label{rmslarge}
\end{equation} 
In the limit $\bar\gamma/\epsilon << 1$, this expression
coincides with the approximate value of eq. (\ref{rmssmall}).
In the opposite limit, $\bar\gamma/\epsilon >>1$, it tends to
$\pi/(2\sqrt 3)$ ($\approx 52\arcdeg$), which corresponds 
to a uniform probability distribution for rotations between 
0 and $\pi/2$, as expected since in this limit the apparent 
rotation is given by $\psi=\phi$. 

The result of the numerical evaluation of eq.
(\ref{rmslarge}) is plotted in Figure 2. 
It indicates that $\sigma_\psi=0.35$ ($20\arcdeg$) 
when $\bar\gamma/\epsilon \approx 0.3$,
which if $\epsilon\approx 0.15$ implies 
\begin{equation}
\bar\gamma\approx 0.05 \quad .
\label{rms}
\end{equation}

The statistical significance of the limited sample
analysed, as well as the crude approach taken towards 
the redshift and ellipticity dependence of the effect,
make this result only an approximate estimate.
It seriously indicates, however, that the observed
departures of the direction of intrinsic polarization
away from the perpendicular to the source axis in
distant radio sources are consistent with the hypothesis
that their origin is the effect of shear with rms value
around or below 5\%. Much larger values of 
$\bar\gamma$ would actually conflict with the observed 
correlation between source polarization and orientation.

Other accurate measurements of
the polarization properties of distant sources 
come from high resolution measurements 
of highly polarized local emission regions
(\cite{Leahy97,Wardle97}). In these observations, 
a very tight correlation is found 
between the plane of polarization and some structural 
direction (such as the direction of intensity gradients),
in agreement with theoretical expectations.
Shear by weak lensing effects would rotate the observed angle
between  polarization and that locally defined direction
by the rms amount $\Delta\psi\approx\bar\gamma/\sqrt 2$ only.

\section{Conclusions}

Measurement of the rotation of the intrinsic polarization with
respect to the local orientation of each jet segment 
in extended radio jets has
been used to probe the mass distribution in intervening
lensing galaxies (\cite{Kronberg91}, \cite{Kronberg96}). 
In this paper we extended this method 
of gravitational ``alignment-breaking'' 
to the case where the change in the observed polarization
properties of distant radio sources is due to the 
weak lensing effect produced by large scale density inhomogeneities.
Indeed, we argued that cosmic shear can have a significant 
impact upon the observed angle $\chi-\psi$ between the 
direction of linear polarization and some structural 
direction of radio sources. In extended
radio sources (sources with large ellipticities $\epsilon\approx 1$)
we found that weak lensing effects are negligible, since the change
in the relative angle $\chi-\psi$ does not exceed
the magnitud of the cosmic shear. On the other hand,
radio sources with intermediate ellipticities, of the order
of $\epsilon\approx 0.1$, can display a significant apparent
rotation of the plane of polarization relative to their major
axis due to gravitational distortions by large-scale structure.

Our study provides an alternative 
approach to evidence the effects of
cosmic shear coherent over
small angular scales, of the order of the source angular
diameter. This approach is complementary to
methods currently used to detect cosmic shear
coherent over the extension of galactic fields
through the search of correlations in
the ellipticities of the galaxies.

Our main conclusion is that density inhomogeneities
do not rotate the direction of linear polarization,
but since they distort elliptical shapes, the angle
between the direction of linear polarization and the
source major axis observed from Earth differs from
the angle at the source (after subtraction of Faraday
rotation). In the limit $\gamma<<\epsilon$,
where $\gamma$ is the value of the shear
induced by the perturbations between the source 
and the observer, and $\epsilon$ is the source 
ellipticity, the apparent
rotation is of order 
$\Delta\psi\approx {\gamma\over \epsilon}$.
This becomes very interesting in the case of sources
with ellipticity of order $\epsilon\approx 0.1$, 
since it implies, in the presence of a shear
of order 1 \%, a rotation of around 5 degrees,  
already comparable or larger than the typical 
errors of a few degrees in Faraday rotation subtraction. 

We have shown that the observed correlation between
polarization properties and source orientation in
distant radio galaxies and quasars can be used to
either constrain or eventually reveal the presence of
cosmic shear. The number of present observations does
not appear to be large enough to test with sufficient
statistical significance whether the observed departures
of the intrinsic polarization angle away from the direction
perpendicular to the source axis may be ascribed to the
effects of scalar shear. Present
observations show that the largest fraction of 
distant radio sources have integrated linear polarization 
in a direction almost perpendicular to the source axis,
with a dispersion around the mean of about $20\arcdeg$. 
We have argued that current observations are consistent 
with the hypothesis that this dispersion is caused by a 
shear with rms value around 5\%, acting upon a population
of radio sources that has a much smaller intrinsic 
dispersion. A rms shear much larger than 5\% would induce 
rms rotations much larger than $20\arcdeg$, and would thus 
be inconsistent with the observed correlation between 
source polarization and orientation. 
With a larger number of observations available, the cosmic 
shear hypothesis could be tested through statistical 
correlations between the polarization vs. orientation angle 
and the source redshift and ellipticity.

Finally, weak lensing by a cosmological background of gravity 
waves is mostly of academic interest, since their effect does
not accumulate with distance as that of scalar perturbations,
and thus the expected distortions are completely negligible
(of order $10^{-6}$). From an academic standpoint, however,
it is interesting to emphasize the differences between tensor
and scalar distortions. Gravity waves produce no focusing
to linear order in the metric perturbation. They do induce
shear plus rigid rotation of images, and also rotate
the plane of polarization of electromagnetic radiation.

\acknowledgments

This work  was supported by 
CONICET,  Universidad de Buenos Aires, and
by an EEC, DG-XII Grant No. CT94-0004.

\clearpage

\begin{figure}
\begin{center}
\leavevmode
\epsfysize=5.25in
\epsfbox{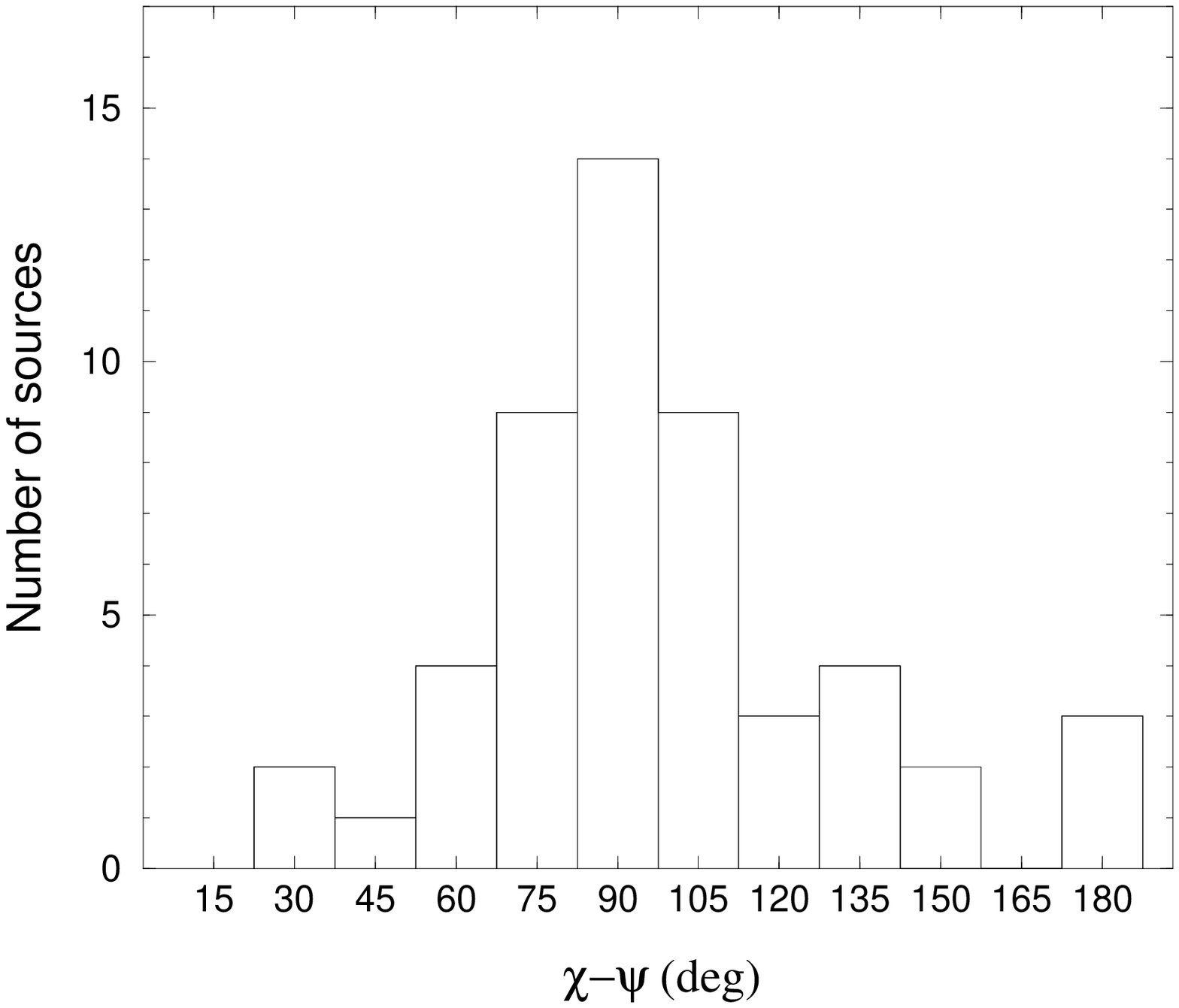}
\caption{Histogram of number of sources vs. $\chi-\psi$ (polarization - 
position angle) for the 51 sources with redshift $z>0.3$ and maximum 
integrated polarization $p_{\rm max}>5\%$}\label{fig1}
\end{center}
\end{figure}

\begin{figure}
\begin{center}
\leavevmode
\epsfysize=5.25in
\epsfbox{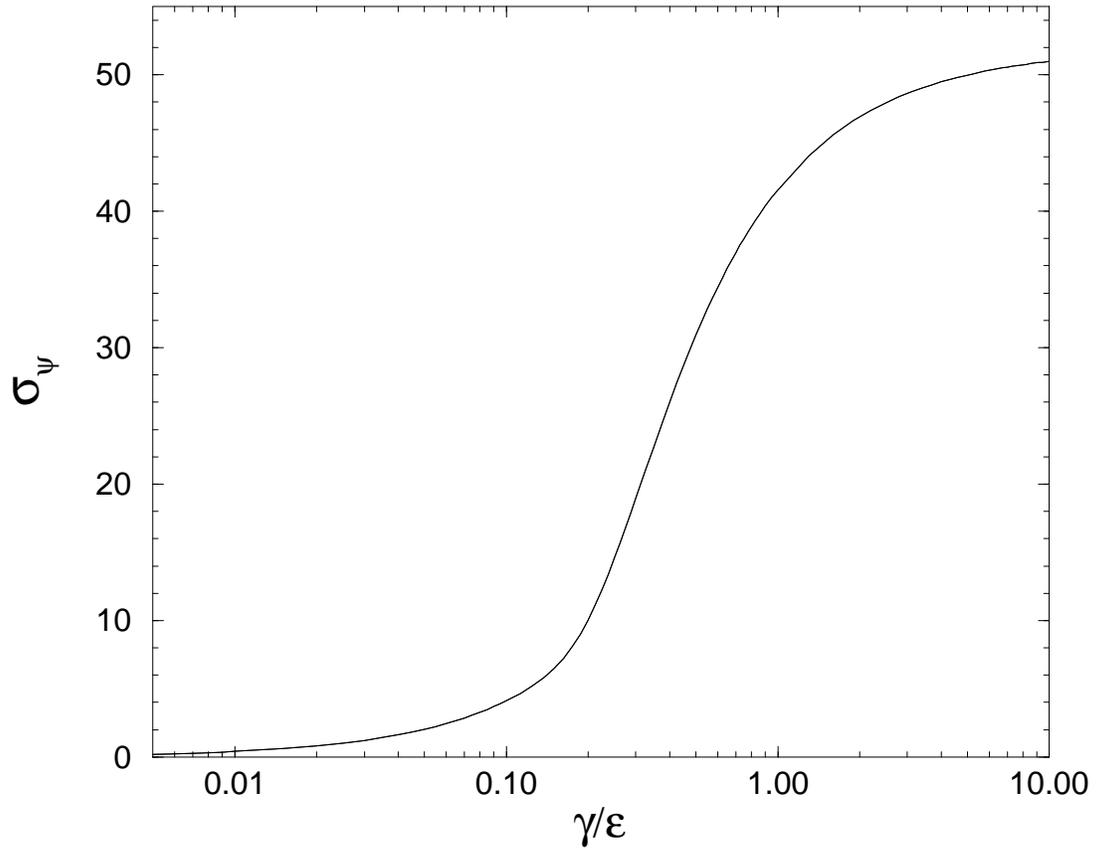}
\caption{Root mean square value of the apparent image rotation
$\sigma_\psi$ induced by a rms shear $\bar\gamma$ upon a source
with ellipticity $\epsilon$, plotted as a function of
$\bar\gamma/\epsilon$}\label{fig2}
\end{center}
\end{figure}

\clearpage

\begin{thebibliography}{}

\bibitem[Abbott \& Wise 1984]{Abbott84}
Abbott, L. F., \& Wise, M. B. 1984, \apj, 282, L47
\bibitem[Balazs 1956]{Balazs56}
Balazs, N. L. 1956, Phys. Rev. 110,236
\bibitem[Bietenholz \& Kronberg 1984]{Bietenholz84} 
Bietenholz, M., \& Kronberg, P. 1984, \apj, 287,L1
\bibitem[Bietenholz 1986]{Bietenholz86} 
Bietenholz, M. 1986, \aj, 91, 1249
\bibitem[Birch 1982]{Birch82}
Birch, P. 1982, \nat, 298, 451
\bibitem[Blandford et al.\ 1991]{Blandford91}
Blandford, R. D., Saust, A. B., Brainerd, T. G., \& Villumsen J. V. 1991,
        \mnras, 251, 600
\bibitem[Bridle, Perley \& Henriksen 1986]{Bridle86}
Bridle, A. H., Perley, R. A., \& Henriksen, R. N. 1986, \aj, 92, 534
\bibitem[Carroll, Field \& Jackiw 1990]{Carroll90}
Carroll, S. M., Field, G. B., \& Jackiw, R. 1990, \prd, 41,1231
\bibitem[Carroll \& Field 1997]{Carroll97}
Carroll, S. M., \& Field, G. B. 1997, preprint astro-ph/9704263
\bibitem[Clarke et al.\ 1980]{Clarke80}
Clarke, J. N., Kronberg, P. P., \& Simard-Normandin, M. 1980, 
        \mnras, 190, 205
\bibitem[Clarke, Norman, \& Burns 1989]{Clarke89}
Clarke, D. A., Norman, M. L., \& Burns, J. O. 1989, \apj, 342, 700
\bibitem[Cooperstock \& Faraoni 1993]{Cooperstock93}
Cooperstock, F.I., \& Faraoni, V. 1993, Class. Quantum Grav.,10,1189 
\bibitem[Dyer \& Shaver 1992]{Dyer92}
Dyer, C. C., \& Shaver, E. G. 1992, \apj, 390, L5
\bibitem[Eisenstein \& Bunn 1997]{Eisenstein97}
Eisenstein, D. J., \& Bunn, E. F. 1997, preprint astro-ph/9704247
\bibitem[Faraoni 1993]{Faraoni93}
Faraoni, V. 1993, \aap, 272, 385 
\bibitem[Gregory et al. 1996]{Gregory96}
Gregory, P.C., Scott, W.K., Douglas, K., \& Condon, J.J. 1996,
ApJ Suppl Ser, 103, 427
\bibitem[Harari \& Sikivie 1992]{Harari92}
Harari, D. D., \& Sikivie, P. 1992, Phys. Lett. B 289, 67
\bibitem[Haves \& Conway 1975]{Haves75}
Haves, P., \& Conway, R. G. 1975, \mnras, 173, 53
\bibitem[Jain \& Seljak 1996]{Jain96}
Jain, B., \& Seljak U. 1996, preprint astro-ph/9611077
\bibitem[Kaiser 1992]{Kaiser92}
Kaiser, N. 1992, \apj, 388, 272
\bibitem[Kaiser 1996]{Kaiser96b}
Kaiser, N. 1996, preprint astro-ph/9610120
\bibitem[Kaiser \& Jaffe 1997]{Kaiser96a}
Kaiser, N., \& Jaffe, A. 1997, \apj, 484, 545
\bibitem[Kendall \& Young 1984]{Kendall84} 
Kendall, D., \& Young, G. A. 1984, \mnras, 207, 637
\bibitem[Killen, Bicknell, \& Ekers 1986]{Killeen86}
Killen, N., Bicknell, G. V., \& Ekers, R. D. 1986, \apj, 302, 306
\bibitem[Kronberg et al. 1991]{Kronberg91}
Kronberg, P. P., Dyer, C. C., Burbidge, E. M., \& Junkkarinen, V. T.
1991, \apj, 367, L1
\bibitem[Kronberg, Dyer, \& R\"{o}ser 1996]{Kronberg96}
Kronberg, P. P., Dyer, C. C., \& R\"{o}ser, H. J. 1996, \apj, 472, 115
\bibitem[Kronberg 1994]{Kronberg94}
Kronberg, P. P. 1994, \nat, 370, 179
\bibitem[Leahy 1997]{Leahy97}
Leahy, J. P. 1997, preprint astro-ph/9704285
\bibitem[Mashhoon 1973]{Mashhoon73}
Mashhoon, B. 1973, \prd,7,2807
\bibitem[Miralda-Escud\'e 1991]{Miralda91}
Miralda-Escud\'e, J. 1991, \apj, 380, 1
\bibitem[Misner, Thorne, \& Wheeler 1973]{MTW}
Misner, C. W., Thorne, K., \& Wheeler, J. A. 1973, Gravitation
(San Francisco: Freeman)
\bibitem[Nodland \& Ralston 1997]{Nodland97}
Nodland, B., \& Ralston, J. P. 1997, \prl, 78,3043 
\bibitem[Perlick \& Hasse 1993]{Perlick93} 
Perlick, V., \& Hasse, W. 1993, Class. Quantum Grav., 10, 147
\bibitem[Pinneault \& Roeder 1977]{Pinneault77}
Pineault, S., \& Roeder, R. C. 1977, \apj, 212, 541
\bibitem[Plebanski 1960]{Plebanski60}
Plebanski, J. 1960, Phys. Rev. 118,1396
\bibitem[Saikia \& Salter 1988]{Saikia88}
Saikia, D. J., \& Salter, C. J. 1988, \araa, 26, 93
\bibitem[Schneider et al.\ 1997]{Schneider97} 
Schneider, P., Van Waerbeke, L., Mellier, Y., Jain, B., Seitz, S. 
        \& Fort, B. 1997, preprint astro-ph/9705122
\bibitem[Seljak 1994]{Seljak94}
Seljak, U. 1994, \apj, 436, 509
\bibitem[Strotskii 1957]{Strotskii57}
Strotskii, G. B. 1957, Sov. Phys. - Doklady 2,226
\bibitem[Su \& Mallett 1980]{Su80}
Su, F. S. O. \& Mallett, R. L. 1980, \apj,238,1111
\bibitem[Tyson et al. 1984]{Valdes84}
Tyson, J. A., Valdes, F., Jarvis, J. F., \& Mills, A. P., Jr. 1984, 
\apj, 281, L59
\bibitem[Tyson, Valdes, \& Wenk 1990]{Tyson90}
Tyson, J. A., Valdes, F., \& Wenk, R. 1990, \apj, 349, L1
\bibitem[Wardle et al. 1997]{Wardle97} 
Wardle, J. F. C., Perley, R. A., \& Cohen, M. H. 1997, 
preprint astro-ph/9705142
\bibitem[Zipoy \& Bertotti 1968]{Bertotti}
Zipoy, D., \& Bertotti, B. 1968, Il Nuovo Cimento, 56B, 195

\end{thebibliography}
\end{document}